\documentclass [%
reprint,
amsmath,amssymb,
aps,
prl,
]{revtex4-1}

\usepackage{graphicx}
\usepackage{amsmath}
\usepackage{amssymb}
\usepackage{bm}
\usepackage{pgf}
\usepackage{color}
\usepackage{amsfonts}
\usepackage{hyperref}
\usepackage{setspace}

\newcommand{\beginsupplement}{%
	\setcounter{table}{0}
	\renewcommand{\thetable}{S\arabic{table}}%
	\setcounter{figure}{0}
	\renewcommand{\thefigure}{S\arabic{figure}}%
	\onecolumngrid
}

\begin{document}

\title{Architected lattices for simultaneous broadband attenuation of airborne sound and mechanical vibrations in all directions}

\author{Osama R. Bilal$^{1,2}$, David Ballagi$^1$, Chiara Daraio$^{2}$}

\affiliation{$^{1}$Department of Mechanical and Process engineering, ETH Zurich, 8092 Zurich, Switzerland}
\affiliation{$^{2}$Division of Engineering and Applied Science, California Institute of Technology, Pasadena, California 91125, USA}

\begin{abstract}
Phononic crystals and acoustic metamaterials are architected lattices designed to control the propagation of acoustic or elastic waves. In these materials, the dispersion properties and the energy transfer are controlled by selecting the lattices' geometry and their constitutive material properties. Most designs, however, only affect one mode of energy propagation, transmitted either as acoustic, airborne sound or as elastic, structural vibrations. Here, we present a design methodology to attenuate both acoustic and elastic waves simultaneously in all polarizations. We experimentally realize the first three-dimensional, load bearing, architected lattice, composed of a single-material, that responds in a broadband frequency range in all directions.
\end{abstract}

\maketitle

Architected materials have the ability to influence the propagation of lattice vibrations or pressure waves across scales. These materials can attenuate  elastic or acoustic energy by supporting the formation of forbidden frequency bands (band gaps) in their dispersion relation, where waves can not propagate. These gaps form through two main mechanisms\cite{deymier2013acoustic}: (1) Bragg scattering, where periodically repeated unitcells scatter waves with wavelength at the same order of the lattice spatial periodicity\cite{kushwaha1993acoustic,sigalas1993band}. (2) Resonances, where locally resonating elements can attenuate waves with wavelength much larger than the lattice periodicity\cite{liu2000locally}. The resonances enable these lattices to retain properties that do not exist in conventional materials, like negative effective mass or stiffness\cite{christensen2015vibrant,cummer2016controlling,ma2016acoustic}. The existence of such band gaps within the frequency spectrum can be utilized for many applications, such as seismic protection\cite{kim2012seismic,brule2014experiments}, vibration or sound insulation\cite{yang2010acoustic,mei2012dark,ma2015purely}, frequency filtering\cite{pennec2004tunable,rupp2010switchable} and wave-guiding\cite{Torres_1999,rupp2007design}, among others\cite{maldovan2013sound}.

Architected lattices can be divided into two broad categories based on the host medium where waves propagate\cite{ cummer2016controlling,christensen2015vibrant}: (i) Acoustic lattices, controlling the propagation of pressure waves in fluids, such as air and water. They usually feature rigid scatterers such as cylinders or spheres, capitalizing on destructive interferences (Bragg-type scattering)\cite{sigalas1993band, kushwaha1996giant, kushwaha1997stop, sanchez1998sound}. Some designs also use resonances (and give rise to negative effective properties): For example, including heavy masses with rubber coating that induce Mie type resonances\cite{liu2000locally}. Other realizations include Helmholtz resonators featuring fluid chambers with single or multiple openings\cite{fang2006ultrasonic,lee2009acoustic} or coiled space\cite{xie2013measurement}. (ii) Elastic lattices, controlling the propagation of stress waves and vibrations in solid materials. They can feature alternating material phases within the unit cell\cite{sigalas1992elastic, kushwaha1993acoustic, kushwaha1994theory, economou1994stop, vasseur1994complete, de1998ultrasonic}, with vast difference in mechanical properties, or single material with geometric features, such as holes\cite{maldovan2009periodic,Bilal_PRE_2011}, leading to Bragg scattering. Another way to attenuate elastic waves is through resonating inclusions, such as pillars or heavy masses\cite{pennec2008low,wu2008evidence,matlack2016composite}. These realizations rely on strong resonance cutting through the dispersion curves to open subwavelength band gaps. Most recent studies on acoustic and elastic metamaterials have focused on the design and characterization of lattices with ever broader (and lower) frequency band gaps, in each separate domain of wave transport\cite{hussein2014dynamics,cummer2016controlling,d2016modeling,matlack2016composite,lucklum2016band, warmuth2017single,d20183d,taniker2015design,wormser2017design}.  

\begin{figure}[b]
	\begin{center}
		\includegraphics[scale = 1]{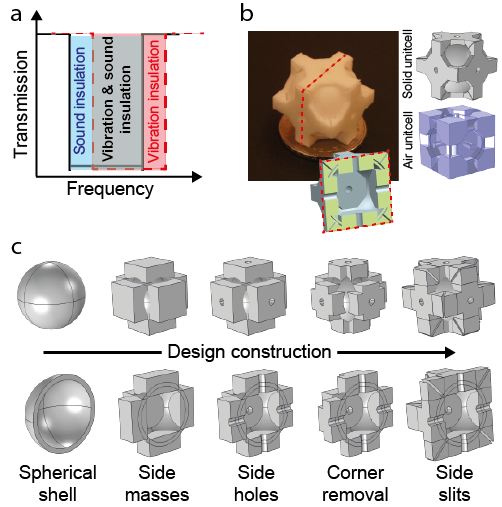}
	\end{center}
	\caption{Color online. (a) Conceptual frequency spectrum of a metamaterial with simultaneous band gaps for airborne sound and mechanical vibrations. (b) Basic building block for a material stopping both sound and vibrations in the same frequency range. Both the solid part and the air within the unit cell is plotted next to the physical prototype. (c) Design sequence starting from a hollow elastic sphere (Top row) Design evolution of the 3D cell, (Bottom row) view of the corresponding mid-section plane cut-out. }
	\label{fig:concept}
\end{figure}

\begin{figure}[t]
	\begin{center}
		\includegraphics[scale =1]{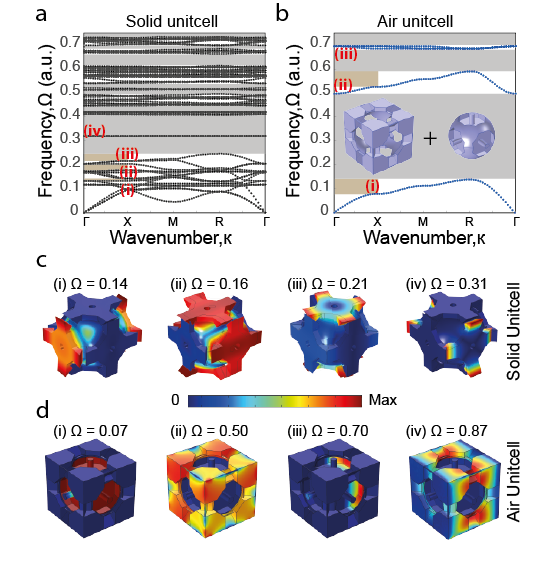}
	\end{center}
	\caption{Color online. Dispersion curves of the metamaterial for (a) mechanical vibrations (b) airborne sound. Full band gaps are highlighted in gray and partial ones are in brown. (c) Selected elastic mode shapes of the solid unit cell. (d) Selected acoustic mode shapes of the air unit cell.}
	\label{fig:unitcell}
\end{figure}

An architected lattice with the ability to attenuate \textit{both} elastic and acoustic waves simultaneously in all directions remains elusive. Such material can be useful for many applications. For example, in airplanes, ships or submarines, when an engine is producing both mechanical vibration and acoustic noise, compromising operational comfort and functionality. In this work, we realize three-dimensional architected lattices that can simultaneously attenuate both acoustic (airborne sound) and elastic waves (vibrations) in all directions, over a broad range of frequencies (Fig. \ref{fig:concept}a-b). Our design methodology capitalizes on both scattering and resonances to open band gaps for sound and vibrations. 

To construct our cubic unitcell, we start with an elastic spherical shell that works as an acoustic chamber resembling a Helmholtz resonator (Fig. \ref{fig:concept}c). The shell also works as an elastic spring connecting six rectangular masses positioned at the center of each of the unitcell faces. This spring-mass arrangement gives rise to Bragg scattering for elastic waves. For the chamber to function as a resonator for acoustic waves, we add a narrow cylindrical channel at each of the unitcell faces. Afterwards, we remove the corner of each face-masses to add an extra acoustic chamber at the eight corners of the unitcell. Finally, we add four resonating ``arms" to each of the six rectangular masses. The added arms function as locally resonating elements for elastic waves, while keeping each face separated from the neighboring faces. The arms also create a narrow slit connecting the corner chambers and introduce a second control over resonances for acoustic waves. 

Based on this design methodology, the position of the band gaps for either sound or vibrations can be easily tuned. For example, by changing the narrow channel radius or the shell thickness. With this method, the attenuated bandwidth of sound frequencies can be chosen independently from the attenuated vibration frequency-ranges. In other words, one can create multiple band gaps in the audible regime for sound waves and have other band gaps in similar (or different) frequencies for elastic vibrations. The realized lattice is load bearing (see Supplementary Information) and the underlying principle of wave attenuation is scale and material agnostic.

\begin{figure*}[!t]
	\begin{center}
		\includegraphics[width = \textwidth]{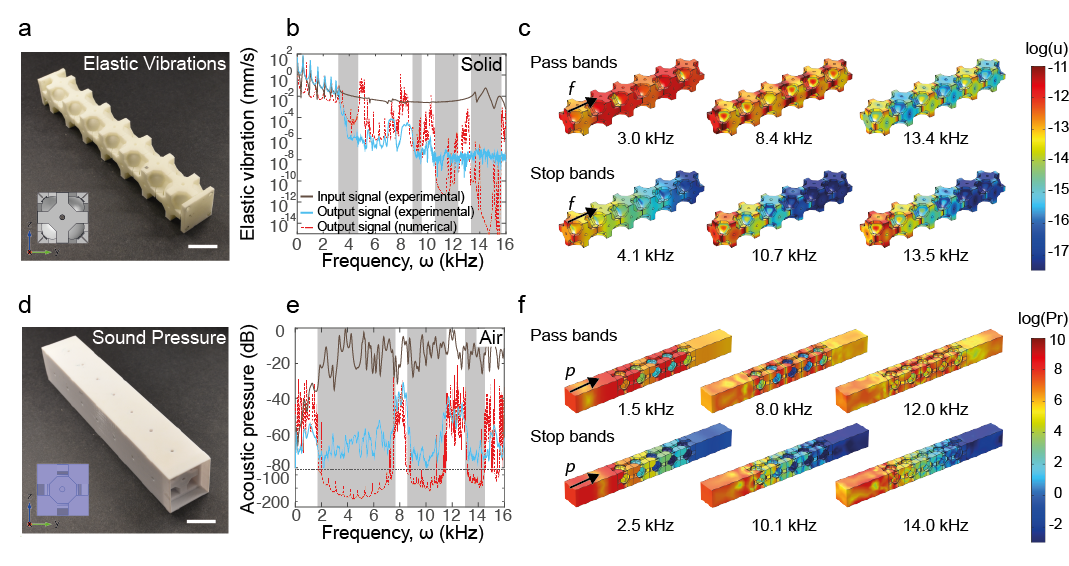}
	\end{center}
	\caption{Color online. (a) Metamaterial used for the elastic vibration experiment. (b) Numerical and experimental frequency response functions for elastic vibrations. The gray shaded areas represent the location of the band gaps calculated with Bloch analysis. (c) Selected mode shapes of the metamaterial at pass bands and stop bands for elastic waves. (d) The same metamaterial enclosed in a tube for sound transmission experiments. (e) Numerical and experimental frequency response functions. (f) Selected mode shapes of the metamaterial within pass and stop bands for sound waves. The scale bars in a and d are 25 mm.}
	\label{fig:sound_and_vibrations}
\end{figure*}

To investigate the validity of our approach, we first consider an infinite medium model, where a single unit cell is analyzed using Bloch periodic boundary conditions\cite{bloch1929quantenmechanik}. We assume small deformations and therefore neglect acousto-elastic coupling. The dispersion curves of the unit cells are calculated using the wave equations for heterogeneous media \cite{graff2012wave} within an infinite medium. We solve both the acoustic and the elastic equations using the finite element method (COMSOL 5.2). The solution is the wavefunction $u (x, \kappa; t) = \tilde{u} (x) \exp^{ (i (\kappa^\intercal x-\omega t))}$, where $\tilde u$ is the Bloch displacement vector, $x$ is the position vector, $\kappa$ is the wavenumber, $\omega$ is frequency and $t$ is time. The dispersion curves relating the wavenumber to the frequency, in non-dimensional units, show band gaps (gray shaded regions) for both elastic (Fig. \ref{fig:unitcell}a) and acoustic waves (Fig. \ref{fig:unitcell}b). The dispersion curves are normalized by multiplying the operational frequency by the unit cell size divided by the speed of the wave in the medium, $\Omega = f a/c$. It should be noted that while the unit cell size is the same in both elastic and acoustic cases, the wave speeds are not. Therefore, having a band gap in both plots, at $\Omega =  0.3$ for instance, does not necessarily guarantee a simultaneous band gap in the dimensional frequency domain. 

To visualize the vibrational mode shapes of the solid unitcell, we superimpose the displacements profiles as a heat map over its geometry for four different frequencies in figure (\ref{fig:unitcell}c). The mode shapes resemble (i) longitudinal, (ii) shear and (iii) rotational modes of the face-masses in the unit cell. We also plot a resonant mode shape of the arms (iv) which manifests itself within the first full band gap in the frequency spectrum at $\Omega$ = 0.31. The acoustic pressure profiles of the air unitcell are superimposed as a heat map over its geometry for four different frequencies in figure (\ref{fig:unitcell}d). The mode shapes show the resonance mode of the  spherical chamber (i and iii) and the corner chambers (ii and iv).

As a proof of concept demonstration, we first realize an array of seven unit cells tessellated along one direction (Fig. \ref{fig:sound_and_vibrations}a,d). We fabricate our samples by additive manufacturing (laser sintering) using polyamide-12 polymer (the measured Young's modulus, through a compression test, and density are $E$ = 0.5 GPa, $\rho$ = 1200 Kg/$m^{3}$). The lattice spacing is \textit{a} = 34 mm.  The elastic response of the metamaterial is characterized by harmonically exciting one of its ends with a mechanical shaker (Br\"uel \& Kjaer Type 4810) and  measuring the transmitted vibrations with a laser Doppler vibrometer LDV (Polytec OFV-505 with a OFV-5000 decoder, using a VD-06 decoder card) at its other end. We sweep through frequencies ranging from 1-16 kHz and record the amplitude of the transmitted vibrations (Fig. \ref{fig:sound_and_vibrations}b). We replicate the experiment numerically using the finite element method, by applying a harmonic load along the x-direction and recording the amplitude of the displacement at the opposite end of the structure. The theoretically predicted band gaps are highlighted in gray in (Fig. \ref{fig:sound_and_vibrations}b). The numerically computed displacements are superimposed as a heat map over the structure for six different frequencies within both pass (top) and stop (bottom) bands in Fig. \ref{fig:sound_and_vibrations}c. Experiments and numerical results agree well. We note the existence of low amplitude regions within the transmission plot that do not coincide with a band gap. They correspond to a pass band with rotational or shear polarizations and can not be excited longitudinally, both in experiments and simulations.

\begin{figure*}[t]
	\begin{center}
		\includegraphics[width = \textwidth]{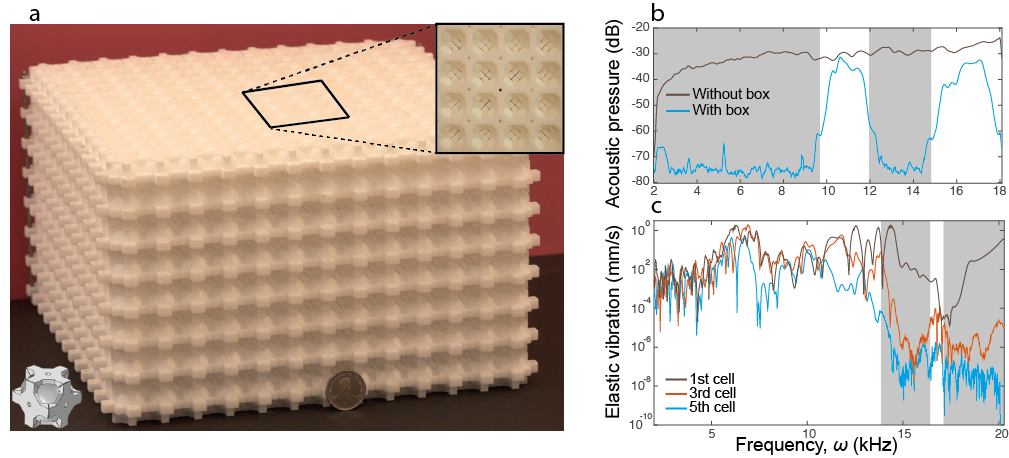}
	\end{center}
	\caption{Color online. (a) A three-dimensional realization of the metamaterial consisting of 13 x 13 x 8 unit cells with a lattice constant a = 25 mm. The material box encloses a piezoelectric transducer for generating mechanical vibrations and a loud speaker for airborne sound. (b) Acoustic frequency response of the metamaterial using a microphone 8 cm above the box  compared to the transmission of same speaker and microphone without our material. (c) Elastic frequency response of the metamaterial using LDV at different distances from the mechanical wave source (1,3,5 unit cells).}
	\label{fig:3D_attenuation}
\end{figure*}

To test the acoustic response of the metamaterial, we enclose the sample within a custom made impedance tube (Fig. \ref{fig:sound_and_vibrations}d) inside an acoustic chamber. Chirp signals are generated with a loud speaker (model Clarion SRE212H) on one end of the tube. Two microphones (G.R.A.S. 40BD) are used to record the generated and transmitted signal, on each side of the metamaterial. We experimentally observe more than 35(dB) reduction in the transmitted  sound along the propagation direction. As a control, we also measure the sound attenuation through a slab of a polyamid-12 solid. We observe that the structured materials, with a density 6 times lower, outperformed the solid barrier by up to 35 (dB) within the band gap frequency range. It is worth noting that our metamaterial is porous. Adding an air opening within the reference solid material dramatically reduces its sound shielding effectiveness due to impedance matching with surrounding air. The difference in attenuation between the control sample (with an air opening) and our structured material exceeds 80 (dB). We model the experiment, considering the tube as a rigid boundary and reproducing the sample's geometry using the finite element method. We generate the excitation as a point source on one of the ends of the tube and plot the intensity of the pressure field on the other end (Fig. \ref{fig:sound_and_vibrations}e). As for mechanical vibrations, the band gaps calculated with Bloch analysis are highlighted in gray. The numerically computed pressure fields are plotted as heat maps for six different frequencies within both pass (top) and stop (bottom) bands in \ref{fig:sound_and_vibrations}f). A good agreement between theory, numerical simulations and experiments is observed.  The results demonstrate the ability of our metamaterial to simultaneously attenuate both airborne sound and mechanical vibrations, in selected frequency ranges. It should be noted that the presence of simultaneous band gaps for elastic vibrations and airborne sound is not automatically granted and has to be designed for. For instance, at 2 kHz, the metamaterial can attenuate sound, but not vibrations. The opposite is true at 12 kHz, where the metamaterial can shield elastic vibrations but not airborne sound.

\begin{figure}[b!]
	\begin{center}
		\includegraphics[width = \textwidth]{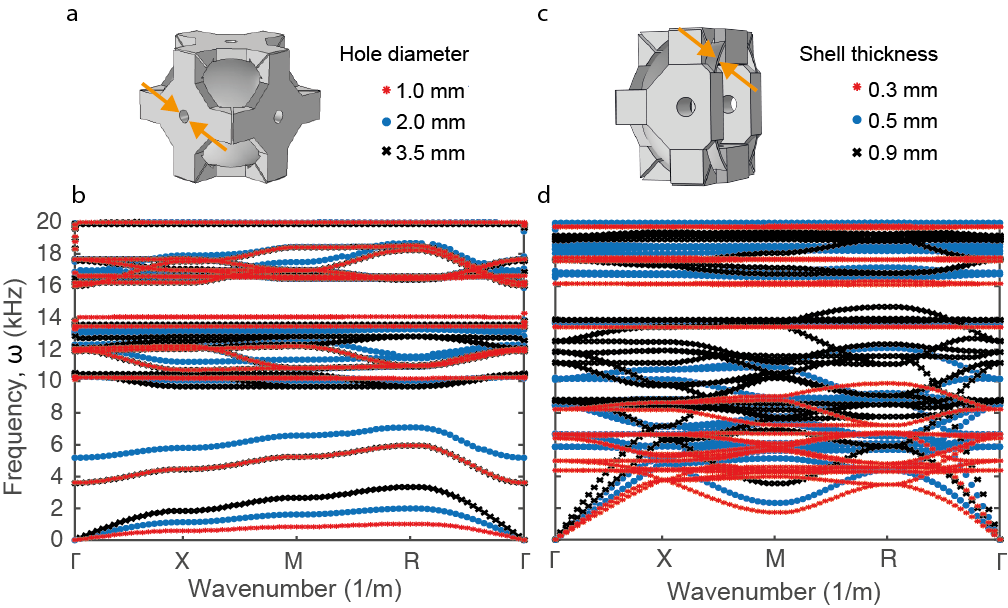}
	\end{center}
	\linespread{1.5}\caption{Color online. Control parameters for changing the position of the band gap for (a-b)acoustic or (c-d)elastic waves independently: (a) Varying  the hole diameter changes the channel width for sound waves and therefore their characteristic acoustic dispersion. (b) Acoustic wave dispersion curves for three different diameters 1, 2 and 3.5 mm. (c) Varying the shell thickness changes the effective coupling between unit cell parts for elastic waves and therefore their characteristic elastic dispersion. (b) Elastic wave dispersion curves for three different shell thicknesses 0.3, 0.5 and 0.9 mm.} 
	\label{fig:Parameters}
\end{figure}

The position of the band gaps within the frequency spectrum, for either sound or vibrations, can be altered by different means. For example, changing the lattice constant would change the band gap position for both sound and vibrations. Using a different constitutive material, for example with higher Young's modulus, would shift the elastic band gaps to higher frequencies, while keeping the acoustic gaps unchanged. However, even while keeping the same material and lattice constant, our design principal allows for decoupling the position of the band gaps for sound and vibrations. For example, consider a unit cell fabricated with polyamide-12 polymer, and a lattice spacing of 25 mm. Changing the radius of the hole at the center of the unit cell (Fig. \ref{fig:Parameters}a), from 0.5 mm to 1.75 mm can change the lower edge of the first complete acoustic band gap from 500 Hz to 2500 Hz (Fig. \ref{fig:Parameters}b), with negligible effects on the elastic waves traveling through the media. Changing the thickness of the shells enclosing the air chamber in each unit cell (Fig. \ref{fig:Parameters}c), from 0.3 mm to 0.9 mm, can shift the lower edge of the elastic band gap from 10 kHz to 15 kHz (Fig. \ref{fig:Parameters}d). The change in shell thickness has a negligible effect on the acoustic response of the metamaterial. Similarly, changing the side openings or the shape/mass of the outer face of the unit cell would significantly change either the acoustic or the elastic response of the metamaterial, respectively, without significantly affecting the other.

To demonstrate the effectiveness of the design methodology in attenuating both sound and vibrations in all directions, we realize a $13 \times 13 \times 8$ lattice with $a = 25$ mm (Fig. \ref{fig:3D_attenuation}a). The fabricated box has a cavity of $3 \times 3 \times 3$ unit cells in its bottom center to host a mechanical transducer and a loud speaker for exciting both structural and sound waves, respectively. Therefore, the effective number of unit cells in any direction ($\pm x, \pm y, +z$) is five. To test the acoustic insulation, we embed a loud speaker (model Clarion SRE212H) inside the metamaterial and measure the amplitude of transmitted sound through the lattice with a 1/4 inch (6.35 mm) microphone (G.R.A.S. 40BD). The test is carried out in an insulated acoustic chamber, moving the microphone in different locations around the metamaterial (see Supplementary Information). We compare the signal propagating through the metamaterials to sound waves recorded without the lattice at the same distance from the source. The measured band gaps span frequencies from 2-9.5 kHz and from 12-14.8 kHz. The attenuated frequency ranges translate to $\approx$ 60\% of the entire audible range. With the metamaterial, we measure more than 35 dB attenuation of the sound wave amplitude in all directions. To test the insulation from mechanical vibrations, we embed a piezoelectric transducer (Piezo Systems  25 x 25 x 2 mm)[www.Piezo.com] within the metamaterial cavity and induce harmonic excitations at different frequencies. We measure the transmitted vibrations through the material on its outer surface at the distance of one, three and five unit cells from the vibration source using the LDV. At the targeted frequency range (highlighted in gray), we observe a significant reduction in the measured wave velocities after the third unit cell. 

Our design methodology allows for the independent tuning of band gap frequency ranges for each domain, in a load bearing metamaterial. Our findings open new opportunities to design advanced multi-functional metamaterials, for application in transportation vessels, machinery and building acoustics.




%


\begin{thebibliography}{42}%
	\makeatletter
	\providecommand \@ifxundefined [1]{%
		\@ifx{#1\undefined}
	}%
	\providecommand \@ifnum [1]{%
		\ifnum #1\expandafter \@firstoftwo
		\else \expandafter \@secondoftwo
		\fi
	}%
	\providecommand \@ifx [1]{%
		\ifx #1\expandafter \@firstoftwo
		\else \expandafter \@secondoftwo
		\fi
	}%
	\providecommand \natexlab [1]{#1}%
	\providecommand \enquote  [1]{``#1''}%
	\providecommand \bibnamefont  [1]{#1}%
	\providecommand \bibfnamefont [1]{#1}%
	\providecommand \citenamefont [1]{#1}%
	\providecommand \href@noop [0]{\@secondoftwo}%
	\providecommand \href [0]{\begingroup \@sanitize@url \@href}%
	\providecommand \@href[1]{\@@startlink{#1}\@@href}%
	\providecommand \@@href[1]{\endgroup#1\@@endlink}%
	\providecommand \@sanitize@url [0]{\catcode `\\12\catcode `\$12\catcode
		`\&12\catcode `\#12\catcode `\^12\catcode `\_12\catcode `\%12\relax}%
	\providecommand \@@startlink[1]{}%
	\providecommand \@@endlink[0]{}%
	\providecommand \url  [0]{\begingroup\@sanitize@url \@url }%
	\providecommand \@url [1]{\endgroup\@href {#1}{\urlprefix }}%
	\providecommand \urlprefix  [0]{URL }%
	\providecommand \Eprint [0]{\href }%
	\providecommand \doibase [0]{http://dx.doi.org/}%
	\providecommand \selectlanguage [0]{\@gobble}%
	\providecommand \bibinfo  [0]{\@secondoftwo}%
	\providecommand \bibfield  [0]{\@secondoftwo}%
	\providecommand \translation [1]{[#1]}%
	\providecommand \BibitemOpen [0]{}%
	\providecommand \bibitemStop [0]{}%
	\providecommand \bibitemNoStop [0]{.\EOS\space}%
	\providecommand \EOS [0]{\spacefactor3000\relax}%
	\providecommand \BibitemShut  [1]{\csname bibitem#1\endcsname}%
	\let\auto@bib@innerbib\@empty
	\bibitem [{\citenamefont {Deymier}(2013)}]{deymier2013acoustic}%
	\BibitemOpen
	\bibfield  {author} {\bibinfo {author} {\bibfnamefont {P.~A.}\ \bibnamefont
			{Deymier}},\ }\href@noop {} {\emph {\bibinfo {title} {Acoustic metamaterials
				and phononic crystals}}},\ Vol.\ \bibinfo {volume} {173}\ (\bibinfo
	{publisher} {Springer Science \& Business Media},\ \bibinfo {year}
	{2013})\BibitemShut {NoStop}%
	\bibitem [{\citenamefont {Kushwaha}\ \emph {et~al.}(1993)\citenamefont
		{Kushwaha}, \citenamefont {Halevi}, \citenamefont {Dobrzynski},\ and\
		\citenamefont {Djafari-Rouhani}}]{kushwaha1993acoustic}%
	\BibitemOpen
	\bibfield  {author} {\bibinfo {author} {\bibfnamefont {M.}~\bibnamefont
			{Kushwaha}}, \bibinfo {author} {\bibfnamefont {P.}~\bibnamefont {Halevi}},
		\bibinfo {author} {\bibfnamefont {L.}~\bibnamefont {Dobrzynski}}, \ and\
		\bibinfo {author} {\bibfnamefont {B.}~\bibnamefont {Djafari-Rouhani}},\
	}\href@noop {} {\bibfield  {journal} {\bibinfo  {journal} {Physical Review
			Letters}\ }\textbf {\bibinfo {volume} {71}},\ \bibinfo {pages} {2022}
	(\bibinfo {year} {1993})}\BibitemShut {NoStop}%
\bibitem [{\citenamefont {Sigalas}\ and\ \citenamefont
	{Economou}(1993)}]{sigalas1993band}%
\BibitemOpen
\bibfield  {author} {\bibinfo {author} {\bibfnamefont {M.}~\bibnamefont
		{Sigalas}}\ and\ \bibinfo {author} {\bibfnamefont {E.}~\bibnamefont
		{Economou}},\ }\href@noop {} {\bibfield  {journal} {\bibinfo  {journal}
		{Solid State Communications}\ }\textbf {\bibinfo {volume} {86}},\ \bibinfo
	{pages} {141} (\bibinfo {year} {1993})}\BibitemShut {NoStop}%
\bibitem [{\citenamefont {Liu}\ \emph {et~al.}(2000)\citenamefont {Liu},
	\citenamefont {Zhang}, \citenamefont {Mao}, \citenamefont {Zhu},
	\citenamefont {Yang}, \citenamefont {Chan},\ and\ \citenamefont
	{Sheng}}]{liu2000locally}%
\BibitemOpen
\bibfield  {author} {\bibinfo {author} {\bibfnamefont {Z.}~\bibnamefont
		{Liu}}, \bibinfo {author} {\bibfnamefont {X.}~\bibnamefont {Zhang}}, \bibinfo
	{author} {\bibfnamefont {Y.}~\bibnamefont {Mao}}, \bibinfo {author}
	{\bibfnamefont {Y.}~\bibnamefont {Zhu}}, \bibinfo {author} {\bibfnamefont
		{Z.}~\bibnamefont {Yang}}, \bibinfo {author} {\bibfnamefont {C.}~\bibnamefont
		{Chan}}, \ and\ \bibinfo {author} {\bibfnamefont {P.}~\bibnamefont {Sheng}},\
}\href@noop {} {\bibfield  {journal} {\bibinfo  {journal} {Science}\ }\textbf
{\bibinfo {volume} {289}},\ \bibinfo {pages} {1734} (\bibinfo {year}
{2000})}\BibitemShut {NoStop}%
\bibitem [{\citenamefont {Christensen}\ \emph {et~al.}(2015)\citenamefont
	{Christensen}, \citenamefont {Kadic}, \citenamefont {Kraft},\ and\
	\citenamefont {Wegener}}]{christensen2015vibrant}%
\BibitemOpen
\bibfield  {author} {\bibinfo {author} {\bibfnamefont {J.}~\bibnamefont
		{Christensen}}, \bibinfo {author} {\bibfnamefont {M.}~\bibnamefont {Kadic}},
	\bibinfo {author} {\bibfnamefont {O.}~\bibnamefont {Kraft}}, \ and\ \bibinfo
	{author} {\bibfnamefont {M.}~\bibnamefont {Wegener}},\ }\href@noop {}
{\bibfield  {journal} {\bibinfo  {journal} {Mrs Communications}\ }\textbf
	{\bibinfo {volume} {5}},\ \bibinfo {pages} {453} (\bibinfo {year}
	{2015})}\BibitemShut {NoStop}%
\bibitem [{\citenamefont {Cummer}\ \emph {et~al.}(2016)\citenamefont {Cummer},
	\citenamefont {Christensen},\ and\ \citenamefont
	{Al{\`u}}}]{cummer2016controlling}%
\BibitemOpen
\bibfield  {author} {\bibinfo {author} {\bibfnamefont {S.~A.}\ \bibnamefont
		{Cummer}}, \bibinfo {author} {\bibfnamefont {J.}~\bibnamefont {Christensen}},
	\ and\ \bibinfo {author} {\bibfnamefont {A.}~\bibnamefont {Al{\`u}}},\
}\href@noop {} {\bibfield  {journal} {\bibinfo  {journal} {Nature Reviews
		Materials}\ }\textbf {\bibinfo {volume} {1}},\ \bibinfo {pages} {16001}
(\bibinfo {year} {2016})}\BibitemShut {NoStop}%
\bibitem [{\citenamefont {Ma}\ and\ \citenamefont
	{Sheng}(2016)}]{ma2016acoustic}%
\BibitemOpen
\bibfield  {author} {\bibinfo {author} {\bibfnamefont {G.}~\bibnamefont
		{Ma}}\ and\ \bibinfo {author} {\bibfnamefont {P.}~\bibnamefont {Sheng}},\
}\href@noop {} {\bibfield  {journal} {\bibinfo  {journal} {Science advances}\
}\textbf {\bibinfo {volume} {2}},\ \bibinfo {pages} {e1501595} (\bibinfo
{year} {2016})}\BibitemShut {NoStop}%
\bibitem [{\citenamefont {Kim}\ and\ \citenamefont
	{Das}(2012)}]{kim2012seismic}%
\BibitemOpen
\bibfield  {author} {\bibinfo {author} {\bibfnamefont {S.-H.}\ \bibnamefont
		{Kim}}\ and\ \bibinfo {author} {\bibfnamefont {M.~P.}\ \bibnamefont {Das}},\
}\href@noop {} {\bibfield  {journal} {\bibinfo  {journal} {Modern Physics
		Letters B}\ }\textbf {\bibinfo {volume} {26}},\ \bibinfo {pages} {1250105}
(\bibinfo {year} {2012})}\BibitemShut {NoStop}%
\bibitem [{\citenamefont {Br{\^u}l{\'e}}\ \emph {et~al.}(2014)\citenamefont
	{Br{\^u}l{\'e}}, \citenamefont {Javelaud}, \citenamefont {Enoch},\ and\
	\citenamefont {Guenneau}}]{brule2014experiments}%
\BibitemOpen
\bibfield  {author} {\bibinfo {author} {\bibfnamefont {S.}~\bibnamefont
		{Br{\^u}l{\'e}}}, \bibinfo {author} {\bibfnamefont {E.}~\bibnamefont
		{Javelaud}}, \bibinfo {author} {\bibfnamefont {S.}~\bibnamefont {Enoch}}, \
	and\ \bibinfo {author} {\bibfnamefont {S.}~\bibnamefont {Guenneau}},\
}\href@noop {} {\bibfield  {journal} {\bibinfo  {journal} {Physical review
		letters}\ }\textbf {\bibinfo {volume} {112}},\ \bibinfo {pages} {133901}
(\bibinfo {year} {2014})}\BibitemShut {NoStop}%
\bibitem [{\citenamefont {Yang}\ \emph {et~al.}(2010)\citenamefont {Yang},
	\citenamefont {Dai}, \citenamefont {Chan}, \citenamefont {Ma},\ and\
	\citenamefont {Sheng}}]{yang2010acoustic}%
\BibitemOpen
\bibfield  {author} {\bibinfo {author} {\bibfnamefont {Z.}~\bibnamefont
		{Yang}}, \bibinfo {author} {\bibfnamefont {H.}~\bibnamefont {Dai}}, \bibinfo
	{author} {\bibfnamefont {N.}~\bibnamefont {Chan}}, \bibinfo {author}
	{\bibfnamefont {G.}~\bibnamefont {Ma}}, \ and\ \bibinfo {author}
	{\bibfnamefont {P.}~\bibnamefont {Sheng}},\ }\href@noop {} {\bibfield
	{journal} {\bibinfo  {journal} { Appl. Phys. Lett.}\ }\textbf {\bibinfo
		{volume} {96}},\ \bibinfo {pages} {041906} (\bibinfo {year}
	{2010})}\BibitemShut {NoStop}%
\bibitem [{\citenamefont {Mei}\ \emph {et~al.}(2012)\citenamefont {Mei},
	\citenamefont {Ma}, \citenamefont {Yang}, \citenamefont {Yang}, \citenamefont
	{Wen},\ and\ \citenamefont {Sheng}}]{mei2012dark}%
\BibitemOpen
\bibfield  {author} {\bibinfo {author} {\bibfnamefont {J.}~\bibnamefont
		{Mei}}, \bibinfo {author} {\bibfnamefont {G.}~\bibnamefont {Ma}}, \bibinfo
	{author} {\bibfnamefont {M.}~\bibnamefont {Yang}}, \bibinfo {author}
	{\bibfnamefont {Z.}~\bibnamefont {Yang}}, \bibinfo {author} {\bibfnamefont
		{W.}~\bibnamefont {Wen}}, \ and\ \bibinfo {author} {\bibfnamefont
		{P.}~\bibnamefont {Sheng}},\ }\href@noop {} {\bibfield  {journal} {\bibinfo
		{journal} {Nature communications}\ }\textbf {\bibinfo {volume} {3}},\
	\bibinfo {pages} {756} (\bibinfo {year} {2012})}\BibitemShut {NoStop}%
\bibitem [{\citenamefont {Ma}\ \emph {et~al.}(2015)\citenamefont {Ma},
	\citenamefont {Wu}, \citenamefont {Huang}, \citenamefont {Zhang},\ and\
	\citenamefont {Zhang}}]{ma2015purely}%
\BibitemOpen
\bibfield  {author} {\bibinfo {author} {\bibfnamefont {F.}~\bibnamefont
		{Ma}}, \bibinfo {author} {\bibfnamefont {J.~H.}\ \bibnamefont {Wu}}, \bibinfo
	{author} {\bibfnamefont {M.}~\bibnamefont {Huang}}, \bibinfo {author}
	{\bibfnamefont {W.}~\bibnamefont {Zhang}}, \ and\ \bibinfo {author}
	{\bibfnamefont {S.}~\bibnamefont {Zhang}},\ }\href@noop {} {\bibfield
	{journal} {\bibinfo  {journal} {Journal of Physics D: Applied Physics}\
	}\textbf {\bibinfo {volume} {48}},\ \bibinfo {pages} {175105} (\bibinfo
	{year} {2015})}\BibitemShut {NoStop}%
\bibitem [{\citenamefont {Pennec}\ \emph {et~al.}(2004)\citenamefont {Pennec},
	\citenamefont {Djafari-Rouhani}, \citenamefont {Vasseur}, \citenamefont
	{Khelif},\ and\ \citenamefont {Deymier}}]{pennec2004tunable}%
\BibitemOpen
\bibfield  {author} {\bibinfo {author} {\bibfnamefont {Y.}~\bibnamefont
		{Pennec}}, \bibinfo {author} {\bibfnamefont {B.}~\bibnamefont
		{Djafari-Rouhani}}, \bibinfo {author} {\bibfnamefont {J.}~\bibnamefont
		{Vasseur}}, \bibinfo {author} {\bibfnamefont {A.}~\bibnamefont {Khelif}}, \
	and\ \bibinfo {author} {\bibfnamefont {P.}~\bibnamefont {Deymier}},\
}\href@noop {} {\bibfield  {journal} {\bibinfo  {journal} {Physical Review
		E}\ }\textbf {\bibinfo {volume} {69}},\ \bibinfo {pages} {046608} (\bibinfo
{year} {2004})}\BibitemShut {NoStop}%
\bibitem [{\citenamefont {Rupp}\ \emph {et~al.}(2010)\citenamefont {Rupp},
	\citenamefont {Dunn},\ and\ \citenamefont {Maute}}]{rupp2010switchable}%
\BibitemOpen
\bibfield  {author} {\bibinfo {author} {\bibfnamefont {C.~J.}\ \bibnamefont
		{Rupp}}, \bibinfo {author} {\bibfnamefont {M.~L.}\ \bibnamefont {Dunn}}, \
	and\ \bibinfo {author} {\bibfnamefont {K.}~\bibnamefont {Maute}},\
}\href@noop {} {\bibfield  {journal} {\bibinfo  {journal} {Appl. Phys. Lett.}\ }\textbf {\bibinfo {volume} {96}},\ \bibinfo {pages} {111902}
(\bibinfo {year} {2010})}\BibitemShut {NoStop}%
\bibitem [{\citenamefont {Torres}\ \emph {et~al.}(1999)\citenamefont {Torres},
	\citenamefont {Montero~de Espinosa}, \citenamefont {Garcia-Pablos},\ and\
	\citenamefont {Garcia}}]{Torres_1999}%
\BibitemOpen
\bibfield  {author} {\bibinfo {author} {\bibfnamefont {M.}~\bibnamefont
		{Torres}}, \bibinfo {author} {\bibfnamefont {F.}~\bibnamefont {Montero~de
			Espinosa}}, \bibinfo {author} {\bibfnamefont {D.}~\bibnamefont
		{Garcia-Pablos}}, \ and\ \bibinfo {author} {\bibfnamefont {N.}~\bibnamefont
		{Garcia}},\ }\href@noop {} {\bibfield  {journal} {\bibinfo  {journal}
		{Physical Review Letters}\ }\textbf {\bibinfo {volume} {82}},\ \bibinfo
	{pages} {3054} (\bibinfo {year} {1999})}\BibitemShut {NoStop}%
\bibitem [{\citenamefont {Rupp}\ \emph {et~al.}(2007)\citenamefont {Rupp},
	\citenamefont {Evgrafov}, \citenamefont {Maute},\ and\ \citenamefont
	{Dunn}}]{rupp2007design}%
\BibitemOpen
\bibfield  {author} {\bibinfo {author} {\bibfnamefont {C.~J.}\ \bibnamefont
		{Rupp}}, \bibinfo {author} {\bibfnamefont {A.}~\bibnamefont {Evgrafov}},
	\bibinfo {author} {\bibfnamefont {K.}~\bibnamefont {Maute}}, \ and\ \bibinfo
	{author} {\bibfnamefont {M.~L.}\ \bibnamefont {Dunn}},\ }\href@noop {}
{\bibfield  {journal} {\bibinfo  {journal} {Structural and Multidisciplinary
			Optimization}\ }\textbf {\bibinfo {volume} {34}},\ \bibinfo {pages} {111}
	(\bibinfo {year} {2007})}\BibitemShut {NoStop}%
\bibitem [{\citenamefont {Maldovan}(2013)}]{maldovan2013sound}%
\BibitemOpen
\bibfield  {author} {\bibinfo {author} {\bibfnamefont {M.}~\bibnamefont
		{Maldovan}},\ }\href@noop {} {\bibfield  {journal} {\bibinfo  {journal}
		{Nature}\ }\textbf {\bibinfo {volume} {503}},\ \bibinfo {pages} {209}
	(\bibinfo {year} {2013})}\BibitemShut {NoStop}%
\bibitem [{\citenamefont {Kushwaha}\ and\ \citenamefont
	{Halevi}(1996)}]{kushwaha1996giant}%
\BibitemOpen
\bibfield  {author} {\bibinfo {author} {\bibfnamefont {M.}~\bibnamefont
		{Kushwaha}}\ and\ \bibinfo {author} {\bibfnamefont {P.}~\bibnamefont
		{Halevi}},\ }\href@noop {} {\bibfield  {journal} {\bibinfo  {journal}
		{Appl. Phys. Lett.}\ }\textbf {\bibinfo {volume} {69}},\ \bibinfo
	{pages} {31} (\bibinfo {year} {1996})}\BibitemShut {NoStop}%
\bibitem [{\citenamefont {Kushwaha}\ and\ \citenamefont
	{Halevi}(1997)}]{kushwaha1997stop}%
\BibitemOpen
\bibfield  {author} {\bibinfo {author} {\bibfnamefont {M.~S.}\ \bibnamefont
		{Kushwaha}}\ and\ \bibinfo {author} {\bibfnamefont {P.}~\bibnamefont
		{Halevi}},\ }\href@noop {} {\bibfield  {journal} {\bibinfo  {journal} {The
			Journal of the Acoustical Society of America}\ }\textbf {\bibinfo {volume}
		{101}},\ \bibinfo {pages} {619} (\bibinfo {year} {1997})}\BibitemShut
{NoStop}%
\bibitem [{\citenamefont {S{\'a}nchez-P{\'e}rez}\ \emph
	{et~al.}(1998)\citenamefont {S{\'a}nchez-P{\'e}rez}, \citenamefont
	{Caballero}, \citenamefont {M{\'a}rtinez-Sala}, \citenamefont {Rubio},
	\citenamefont {S{\'a}nchez-Dehesa}, \citenamefont {Meseguer}, \citenamefont
	{Llinares},\ and\ \citenamefont {G{\'a}lvez}}]{sanchez1998sound}%
\BibitemOpen
\bibfield  {author} {\bibinfo {author} {\bibfnamefont {J.~V.}\ \bibnamefont
		{S{\'a}nchez-P{\'e}rez}}, \bibinfo {author} {\bibfnamefont {D.}~\bibnamefont
		{Caballero}}, \bibinfo {author} {\bibfnamefont {R.}~\bibnamefont
		{M{\'a}rtinez-Sala}}, \bibinfo {author} {\bibfnamefont {C.}~\bibnamefont
		{Rubio}}, \bibinfo {author} {\bibfnamefont {J.}~\bibnamefont
		{S{\'a}nchez-Dehesa}}, \bibinfo {author} {\bibfnamefont {F.}~\bibnamefont
		{Meseguer}}, \bibinfo {author} {\bibfnamefont {J.}~\bibnamefont {Llinares}},
	\ and\ \bibinfo {author} {\bibfnamefont {F.}~\bibnamefont {G{\'a}lvez}},\
}\href@noop {} {\bibfield  {journal} {\bibinfo  {journal} {Physical Review
		Letters}\ }\textbf {\bibinfo {volume} {80}},\ \bibinfo {pages} {5325}
(\bibinfo {year} {1998})}\BibitemShut {NoStop}%
\bibitem [{\citenamefont {Fang}\ \emph {et~al.}(2006)\citenamefont {Fang},
	\citenamefont {Xi}, \citenamefont {Xu}, \citenamefont {Ambati}, \citenamefont
	{Srituravanich}, \citenamefont {Sun},\ and\ \citenamefont
	{Zhang}}]{fang2006ultrasonic}%
\BibitemOpen
\bibfield  {author} {\bibinfo {author} {\bibfnamefont {N.}~\bibnamefont
		{Fang}}, \bibinfo {author} {\bibfnamefont {D.}~\bibnamefont {Xi}}, \bibinfo
	{author} {\bibfnamefont {J.}~\bibnamefont {Xu}}, \bibinfo {author}
	{\bibfnamefont {M.}~\bibnamefont {Ambati}}, \bibinfo {author} {\bibfnamefont
		{W.}~\bibnamefont {Srituravanich}}, \bibinfo {author} {\bibfnamefont
		{C.}~\bibnamefont {Sun}}, \ and\ \bibinfo {author} {\bibfnamefont
		{X.}~\bibnamefont {Zhang}},\ }\href@noop {} {\bibfield  {journal} {\bibinfo
		{journal} {Nature materials}\ }\textbf {\bibinfo {volume} {5}},\ \bibinfo
	{pages} {452} (\bibinfo {year} {2006})}\BibitemShut {NoStop}%
\bibitem [{\citenamefont {Lee}\ \emph {et~al.}(2009)\citenamefont {Lee},
	\citenamefont {Park}, \citenamefont {Seo}, \citenamefont {Wang},\ and\
	\citenamefont {Kim}}]{lee2009acoustic}%
\BibitemOpen
\bibfield  {author} {\bibinfo {author} {\bibfnamefont {S.~H.}\ \bibnamefont
		{Lee}}, \bibinfo {author} {\bibfnamefont {C.~M.}\ \bibnamefont {Park}},
	\bibinfo {author} {\bibfnamefont {Y.~M.}\ \bibnamefont {Seo}}, \bibinfo
	{author} {\bibfnamefont {Z.~G.}\ \bibnamefont {Wang}}, \ and\ \bibinfo
	{author} {\bibfnamefont {C.~K.}\ \bibnamefont {Kim}},\ }\href@noop {}
{\bibfield  {journal} {\bibinfo  {journal} {Journal of Physics: Condensed
			Matter}\ }\textbf {\bibinfo {volume} {21}},\ \bibinfo {pages} {175704}
	(\bibinfo {year} {2009})}\BibitemShut {NoStop}%
\bibitem [{\citenamefont {Xie}\ \emph {et~al.}(2013)\citenamefont {Xie},
	\citenamefont {Popa}, \citenamefont {Zigoneanu},\ and\ \citenamefont
	{Cummer}}]{xie2013measurement}%
\BibitemOpen
\bibfield  {author} {\bibinfo {author} {\bibfnamefont {Y.}~\bibnamefont
		{Xie}}, \bibinfo {author} {\bibfnamefont {B.-I.}\ \bibnamefont {Popa}},
	\bibinfo {author} {\bibfnamefont {L.}~\bibnamefont {Zigoneanu}}, \ and\
	\bibinfo {author} {\bibfnamefont {S.~A.}\ \bibnamefont {Cummer}},\
}\href@noop {} {\bibfield  {journal} {\bibinfo  {journal} {Physical review
		letters}\ }\textbf {\bibinfo {volume} {110}},\ \bibinfo {pages} {175501}
(\bibinfo {year} {2013})}\BibitemShut {NoStop}%
\bibitem [{\citenamefont {Sigalas}\ and\ \citenamefont
	{Economou}(1992)}]{sigalas1992elastic}%
\BibitemOpen
\bibfield  {author} {\bibinfo {author} {\bibfnamefont {M.~M.}\ \bibnamefont
		{Sigalas}}\ and\ \bibinfo {author} {\bibfnamefont {E.~N.}\ \bibnamefont
		{Economou}},\ }\href@noop {} {\bibfield  {journal} {\bibinfo  {journal}
		{Journal of Sound Vibration}\ }\textbf {\bibinfo {volume} {158}},\ \bibinfo
	{pages} {377} (\bibinfo {year} {1992})}\BibitemShut {NoStop}%
\bibitem [{\citenamefont {Kushwaha}\ \emph {et~al.}(1994)\citenamefont
	{Kushwaha}, \citenamefont {Halevi}, \citenamefont {Martinez}, \citenamefont
	{Dobrzynski},\ and\ \citenamefont {Djafari-Rouhani}}]{kushwaha1994theory}%
\BibitemOpen
\bibfield  {author} {\bibinfo {author} {\bibfnamefont {M.~S.}\ \bibnamefont
		{Kushwaha}}, \bibinfo {author} {\bibfnamefont {P.}~\bibnamefont {Halevi}},
	\bibinfo {author} {\bibfnamefont {G.}~\bibnamefont {Martinez}}, \bibinfo
	{author} {\bibfnamefont {L.}~\bibnamefont {Dobrzynski}}, \ and\ \bibinfo
	{author} {\bibfnamefont {B.}~\bibnamefont {Djafari-Rouhani}},\ }\href@noop {}
{\bibfield  {journal} {\bibinfo  {journal} {Physical Review B}\ }\textbf
	{\bibinfo {volume} {49}},\ \bibinfo {pages} {2313} (\bibinfo {year}
	{1994})}\BibitemShut {NoStop}%
\bibitem [{\citenamefont {Economou}\ and\ \citenamefont
	{Sigalas}(1994)}]{economou1994stop}%
\BibitemOpen
\bibfield  {author} {\bibinfo {author} {\bibfnamefont {E.}~\bibnamefont
		{Economou}}\ and\ \bibinfo {author} {\bibfnamefont {M.}~\bibnamefont
		{Sigalas}},\ }\href@noop {} {\bibfield  {journal} {\bibinfo  {journal} {The
			Journal of the Acoustical Society of America}\ }\textbf {\bibinfo {volume}
		{95}},\ \bibinfo {pages} {1734} (\bibinfo {year} {1994})}\BibitemShut
{NoStop}%
\bibitem [{\citenamefont {Vasseur}\ \emph {et~al.}(1994)\citenamefont
	{Vasseur}, \citenamefont {Djafari-Rouhani}, \citenamefont {Dobrzynski},
	\citenamefont {Kushwaha},\ and\ \citenamefont
	{Halevi}}]{vasseur1994complete}%
\BibitemOpen
\bibfield  {author} {\bibinfo {author} {\bibfnamefont {J.}~\bibnamefont
		{Vasseur}}, \bibinfo {author} {\bibfnamefont {B.}~\bibnamefont
		{Djafari-Rouhani}}, \bibinfo {author} {\bibfnamefont {L.}~\bibnamefont
		{Dobrzynski}}, \bibinfo {author} {\bibfnamefont {M.}~\bibnamefont
		{Kushwaha}}, \ and\ \bibinfo {author} {\bibfnamefont {P.}~\bibnamefont
		{Halevi}},\ }\href@noop {} {\bibfield  {journal} {\bibinfo  {journal}
		{Journal of Physics: Condensed Matter}\ }\textbf {\bibinfo {volume} {6}},\
	\bibinfo {pages} {8759} (\bibinfo {year} {1994})}\BibitemShut {NoStop}%
\bibitem [{\citenamefont {De~Espinosa}\ \emph {et~al.}(1998)\citenamefont
	{De~Espinosa}, \citenamefont {Jimenez},\ and\ \citenamefont
	{Torres}}]{de1998ultrasonic}%
\BibitemOpen
\bibfield  {author} {\bibinfo {author} {\bibfnamefont {F.~M.}\ \bibnamefont
		{De~Espinosa}}, \bibinfo {author} {\bibfnamefont {E.}~\bibnamefont
		{Jimenez}}, \ and\ \bibinfo {author} {\bibfnamefont {M.}~\bibnamefont
		{Torres}},\ }\href@noop {} {\bibfield  {journal} {\bibinfo  {journal}
		{Physical Review Letters}\ }\textbf {\bibinfo {volume} {80}},\ \bibinfo
	{pages} {1208} (\bibinfo {year} {1998})}\BibitemShut {NoStop}%
\bibitem [{\citenamefont {Maldovan}\ and\ \citenamefont
	{Thomas}(2009)}]{maldovan2009periodic}%
\BibitemOpen
\bibfield  {author} {\bibinfo {author} {\bibfnamefont {M.}~\bibnamefont
		{Maldovan}}\ and\ \bibinfo {author} {\bibfnamefont {E.~L.}\ \bibnamefont
		{Thomas}},\ }\href@noop {} {\emph {\bibinfo {title} {Periodic materials and
			interference lithography: for photonics, phononics and mechanics}}}\
(\bibinfo  {publisher} {John Wiley \& Sons},\ \bibinfo {year}
{2009})\BibitemShut {NoStop}%
\bibitem [{\citenamefont {Bilal}\ and\ \citenamefont
	{Hussein}(2011)}]{Bilal_PRE_2011}%
\BibitemOpen
\bibfield  {author} {\bibinfo {author} {\bibfnamefont {O.~R.}\ \bibnamefont
		{Bilal}}\ and\ \bibinfo {author} {\bibfnamefont {M.~I.}\ \bibnamefont
		{Hussein}},\ }\href@noop {} {\bibfield  {journal} {\bibinfo  {journal}
		{Physical Review E}\ }\textbf {\bibinfo {volume} {84}},\ \bibinfo {pages}
	{065701} (\bibinfo {year} {2011})}\BibitemShut {NoStop}%
\bibitem [{\citenamefont {Pennec}\ \emph {et~al.}(2008)\citenamefont {Pennec},
	\citenamefont {Djafari-Rouhani}, \citenamefont {Larabi}, \citenamefont
	{Vasseur},\ and\ \citenamefont {Hladky-Hennion}}]{pennec2008low}%
\BibitemOpen
\bibfield  {author} {\bibinfo {author} {\bibfnamefont {Y.}~\bibnamefont
		{Pennec}}, \bibinfo {author} {\bibfnamefont {B.}~\bibnamefont
		{Djafari-Rouhani}}, \bibinfo {author} {\bibfnamefont {H.}~\bibnamefont
		{Larabi}}, \bibinfo {author} {\bibfnamefont {J.}~\bibnamefont {Vasseur}}, \
	and\ \bibinfo {author} {\bibfnamefont {A.}~\bibnamefont {Hladky-Hennion}},\
}\href@noop {} {\bibfield  {journal} {\bibinfo  {journal} {Physical Review
		B}\ }\textbf {\bibinfo {volume} {78}},\ \bibinfo {pages} {104105} (\bibinfo
{year} {2008})}\BibitemShut {NoStop}%
\bibitem [{\citenamefont {Wu}\ \emph {et~al.}(2008)\citenamefont {Wu},
	\citenamefont {Huang}, \citenamefont {Tsai},\ and\ \citenamefont
	{Wu}}]{wu2008evidence}%
\BibitemOpen
\bibfield  {author} {\bibinfo {author} {\bibfnamefont {T.-T.}\ \bibnamefont
		{Wu}}, \bibinfo {author} {\bibfnamefont {Z.-G.}\ \bibnamefont {Huang}},
	\bibinfo {author} {\bibfnamefont {T.-C.}\ \bibnamefont {Tsai}}, \ and\
	\bibinfo {author} {\bibfnamefont {T.-C.}\ \bibnamefont {Wu}},\ }\href@noop {}
{\bibfield  {journal} {\bibinfo  {journal} {Appl. Phys. Lett.}\
	}\textbf {\bibinfo {volume} {93}},\ \bibinfo {pages} {111902} (\bibinfo
	{year} {2008})}\BibitemShut {NoStop}%
\bibitem [{\citenamefont {Matlack}\ \emph {et~al.}(2016)\citenamefont
	{Matlack}, \citenamefont {Bauhofer}, \citenamefont {Kr{\"o}del},
	\citenamefont {Palermo},\ and\ \citenamefont
	{Daraio}}]{matlack2016composite}%
\BibitemOpen
\bibfield  {author} {\bibinfo {author} {\bibfnamefont {K.~H.}\ \bibnamefont
		{Matlack}}, \bibinfo {author} {\bibfnamefont {A.}~\bibnamefont {Bauhofer}},
	\bibinfo {author} {\bibfnamefont {S.}~\bibnamefont {Kr{\"o}del}}, \bibinfo
	{author} {\bibfnamefont {A.}~\bibnamefont {Palermo}}, \ and\ \bibinfo
	{author} {\bibfnamefont {C.}~\bibnamefont {Daraio}},\ }\href@noop {}
{\bibfield  {journal} {\bibinfo  {journal} {Proc Natl Acad Sci}\ }\textbf {\bibinfo {volume} {113}},\ \bibinfo {pages}
	{8386} (\bibinfo {year} {2016})}\BibitemShut {NoStop}%
\bibitem [{\citenamefont {Hussein}\ \emph {et~al.}(2014)\citenamefont
	{Hussein}, \citenamefont {Leamy},\ and\ \citenamefont
	{Ruzzene}}]{hussein2014dynamics}%
\BibitemOpen
\bibfield  {author} {\bibinfo {author} {\bibfnamefont {M.~I.}\ \bibnamefont
		{Hussein}}, \bibinfo {author} {\bibfnamefont {M.~J.}\ \bibnamefont {Leamy}},
	\ and\ \bibinfo {author} {\bibfnamefont {M.}~\bibnamefont {Ruzzene}},\
}\href@noop {} {\bibfield  {journal} {\bibinfo  {journal} {Applied Mechanics
		Reviews}\ }\textbf {\bibinfo {volume} {66}},\ \bibinfo {pages} {040802}
(\bibinfo {year} {2014})}\BibitemShut {NoStop}%
\bibitem [{\citenamefont {D'Alessandro}\ \emph {et~al.}(2016)\citenamefont
	{D'Alessandro}, \citenamefont {Belloni}, \citenamefont {Ardito},
	\citenamefont {Corigliano},\ and\ \citenamefont {Braghin}}]{d2016modeling}%
\BibitemOpen
\bibfield  {author} {\bibinfo {author} {\bibfnamefont {L.}~\bibnamefont
		{D'Alessandro}}, \bibinfo {author} {\bibfnamefont {E.}~\bibnamefont
		{Belloni}}, \bibinfo {author} {\bibfnamefont {R.}~\bibnamefont {Ardito}},
	\bibinfo {author} {\bibfnamefont {A.}~\bibnamefont {Corigliano}}, \ and\
	\bibinfo {author} {\bibfnamefont {F.}~\bibnamefont {Braghin}},\ }\href@noop
{} {\bibfield  {journal} {\bibinfo  {journal} {Appl. Phys. Lett.}\
	}\textbf {\bibinfo {volume} {109}},\ \bibinfo {pages} {221907} (\bibinfo
	{year} {2016})}\BibitemShut {NoStop}%
\bibitem [{\citenamefont {Lucklum}\ and\ \citenamefont
	{Vellekoop}()}]{lucklum2016band}%
\BibitemOpen
\bibfield  {author} {\bibinfo {author} {\bibfnamefont {F.}~\bibnamefont
		{Lucklum}}\ and\ \bibinfo {author} {\bibfnamefont {M.~J.}\ \bibnamefont
		{Vellekoop}},\ }in\ \href@noop {} {\emph {\bibinfo {booktitle} {Ultrasonics
			Symposium (IUS), 2016 IEEE International}}}\BibitemShut {NoStop}%
\bibitem [{\citenamefont {Warmuth}\ \emph {et~al.}(2017)\citenamefont
	{Warmuth}, \citenamefont {Wormser},\ and\ \citenamefont
	{K{\"o}rner}}]{warmuth2017single}%
\BibitemOpen
\bibfield  {author} {\bibinfo {author} {\bibfnamefont {F.}~\bibnamefont
		{Warmuth}}, \bibinfo {author} {\bibfnamefont {M.}~\bibnamefont {Wormser}}, \
	and\ \bibinfo {author} {\bibfnamefont {C.}~\bibnamefont {K{\"o}rner}},\
}\href@noop {} {\bibfield  {journal} {\bibinfo  {journal} {Scientific
		reports}\ }\textbf {\bibinfo {volume} {7}},\ \bibinfo {pages} {3843}
(\bibinfo {year} {2017})}\BibitemShut {NoStop}%
\bibitem [{\citenamefont {D’Alessandro}\ \emph {et~al.}(2018)\citenamefont
	{D’Alessandro}, \citenamefont {Zega}, \citenamefont {Ardito},\ and\
	\citenamefont {Corigliano}}]{d20183d}%
\BibitemOpen
\bibfield  {author} {\bibinfo {author} {\bibfnamefont {L.}~\bibnamefont
		{D’Alessandro}}, \bibinfo {author} {\bibfnamefont {V.}~\bibnamefont
		{Zega}}, \bibinfo {author} {\bibfnamefont {R.}~\bibnamefont {Ardito}}, \ and\
	\bibinfo {author} {\bibfnamefont {A.}~\bibnamefont {Corigliano}},\
}\href@noop {} {\bibfield  {journal} {\bibinfo  {journal} {Scientific
		reports}\ }\textbf {\bibinfo {volume} {8}},\ \bibinfo {pages} {2262}
(\bibinfo {year} {2018})}\BibitemShut {NoStop}%
\bibitem [{\citenamefont {Taniker}\ and\ \citenamefont
	{Yilmaz}(2015)}]{taniker2015design}%
\BibitemOpen
\bibfield  {author} {\bibinfo {author} {\bibfnamefont {S.}~\bibnamefont
		{Taniker}}\ and\ \bibinfo {author} {\bibfnamefont {C.}~\bibnamefont
		{Yilmaz}},\ }\href@noop {} {\bibfield  {journal} {\bibinfo  {journal}
		{International Journal of Solids and Structures}\ }\textbf {\bibinfo {volume}
		{72}},\ \bibinfo {pages} {88} (\bibinfo {year} {2015})}\BibitemShut {NoStop}%
\bibitem [{\citenamefont {Wormser}\ \emph {et~al.}(2017)\citenamefont
	{Wormser}, \citenamefont {Wein}, \citenamefont {Stingl},\ and\ \citenamefont
	{K{\"o}rner}}]{wormser2017design}%
\BibitemOpen
\bibfield  {author} {\bibinfo {author} {\bibfnamefont {M.}~\bibnamefont
		{Wormser}}, \bibinfo {author} {\bibfnamefont {F.}~\bibnamefont {Wein}},
	\bibinfo {author} {\bibfnamefont {M.}~\bibnamefont {Stingl}}, \ and\ \bibinfo
	{author} {\bibfnamefont {C.}~\bibnamefont {K{\"o}rner}},\ }\href@noop {}
{\bibfield  {journal} {\bibinfo  {journal} {Materials}\ }\textbf {\bibinfo
		{volume} {10}},\ \bibinfo {pages} {1125} (\bibinfo {year}
	{2017})}\BibitemShut {NoStop}%
\bibitem [{\citenamefont {Bloch}(1929)}]{bloch1929quantenmechanik}%
\BibitemOpen
\bibfield  {author} {\bibinfo {author} {\bibfnamefont {F.}~\bibnamefont
		{Bloch}},\ }\href@noop {} {\bibfield  {journal} {\bibinfo  {journal} {Journal
			of Physics}\ }\textbf {\bibinfo {volume} {52}},\ \bibinfo {pages} {555}
	(\bibinfo {year} {1929})}\BibitemShut {NoStop}%
\bibitem [{\citenamefont {Graff}(2012)}]{graff2012wave}%
\BibitemOpen
\bibfield  {author} {\bibinfo {author} {\bibfnamefont {K.~F.}\ \bibnamefont
		{Graff}},\ }\href@noop {} {\emph {\bibinfo {title} {Wave motion in elastic
			solids}}}\ (\bibinfo  {publisher} {Courier Corporation},\ \bibinfo {year}
{2012})\BibitemShut {NoStop}%
\end{thebibliography}

\beginsupplement 

\begin{figure}[b]
	\begin{center}
		\includegraphics[scale =1]{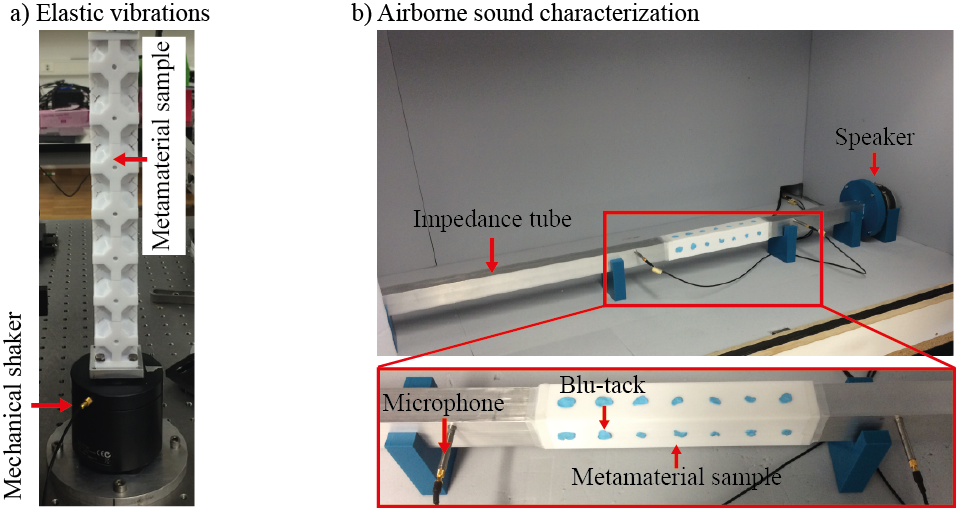}
	\end{center}
	\linespread{1.5}\caption{One-dimensional metamaterial characterization: (a) Mechanical vibration excitation of a 7$\times$1 metamaterial sample using a mechanical shaker. (b) Airborne sound excitation of a 7$\times$1 metamaterial sample using a loud speaker and a custom made impedance tube with a square cross section.}
	\label{fig:Elastic and acoustic}
\end{figure}

\begin{center}
{\Large {\textbf{Supplementary information}:\\~\\ Architected lattices for simultaneous broadband attenuation of airborne sound and mechanical vibrations in all directions}}
\end{center}
\begin{spacing}{1.75}
\paragraph{One-dimensional metamaterial characterization:} To test our metamaterial properties, we fabricate two arrays composed of seven unit cells. For elastic vibration testing, we add two thin plates (5 mm) at each end of the metamaterial array. We mount an electromechanical shaker against one of the plates and measure the transmitted signal at the second plate (Fig.\ref{fig:Elastic and acoustic}a). We cover the free end (i.e., the second plate) with a reflective tape and record its movement (displacement and velocity) using a laser Doppler vibrometer. The experimental setup guarantees complete isolation of the metamaterial sample from any undesired vibrations through the table. The excitation signal is sent to the electromechanical shaker from the PC through an audio amplifier (Topping TP22). The measured velocities are sent back to the PC through a lock-in amplifier  model (Z\"urich Instruments HF2LI).

\begin{figure}
	\begin{center}
		\includegraphics[scale =.125]{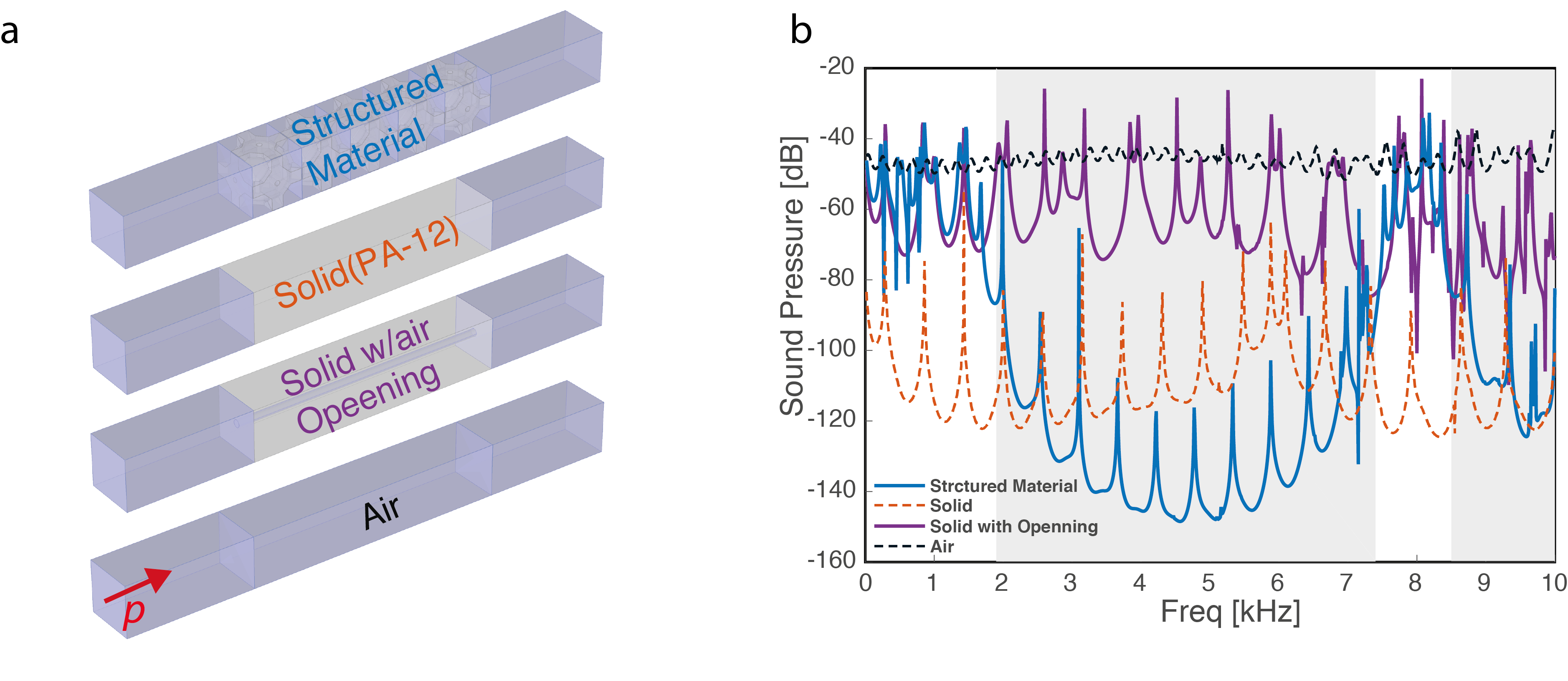}
	\end{center}
	\linespread{1.5}\caption{Comparison of the acoustic attenuation of different materials:(a) Schematic of tested materials (i) structured material with air openings, (ii) air-tight solid block made of PA-12 (iii) Pa-12 block with a cylindrical opening with 5 mm diameter and (iv) an open air channel as a reference. (b) The numerical frequency response function FRF of sound pressure field at the end of the impedance tube.}
	\label{fig:acoustic_comp}
\end{figure}


For airborne acoustic testing, we fabricate the metamaterial enclosed in a tube with a square cross section. The printed tube has circular holes aligned with those in the metamaterial sample, to ease the removal of excess powder from the printing process. The side holes are then sound proofed using Blu-tack, to prevent any sound leak from the metamaterial to the chamber and to ensure full transmission of the wave through the longitudinal direction of the metamaterial. We fit both ends of the metamaterial in a custom impedance tube with a square profile. The tube has two microphones mounted at a distance of 70 mm from the edges of the metamaterial sample. A loud speaker is mounted at one end of the tube, while the other end is fitted within the padding of the acoustic chamber (Fig.\ref{fig:Elastic and acoustic}b).

As a control, we simulate the sound attenuation characteristics of an open air channel, a solid block made of PA-12 and a PA-12 block with a small open cylindrical channel (diameter = 5 mm). We use COMSOL multi-physics acoustics module to simulate the sound pressure fields. An identical impedance tube to the custom made one in Fig. \ref{fig:Elastic and acoustic}b is modeled in the numerical simulations. The pressure wave is introduced at one end of the tube as a point source and the resulting pressure field is measured at the other end of the sample. We compare the performance of the three samples against our structured metamaterial (Fig. \ref{fig:acoustic_comp}a). The open air channel has no attenuation capabilities, as expected, therefore it sets the bar for sound pressure level. Due to impedance mismatch ($Z_{air}$ = 0.000445 kg/m$^{2}$s$ \times$$10^6$, $Z_{PA-12}$ =  1.5  kg/m$^{2}$s$ \times$$10^6$), an air-tight block of PA-12 attenuates a large amount of the incident sound energy, however, the effectiveness of such approach degrades exponentially with a small air-opening. Such scenario is common with assembly of parts, mechanisms or when open air systems are required (e.g., for cooling purposes). In the case of our structured materials, the attenuation level is equivalent to the homogeneous material with an open channel in pass band frequencies. Within the frequency range of the band gap, the structured material has a superior attenuation profile for sound waves compared to the three other modeled samples.

\begin{figure}	
	\begin{center}
		\includegraphics[width = \textwidth]{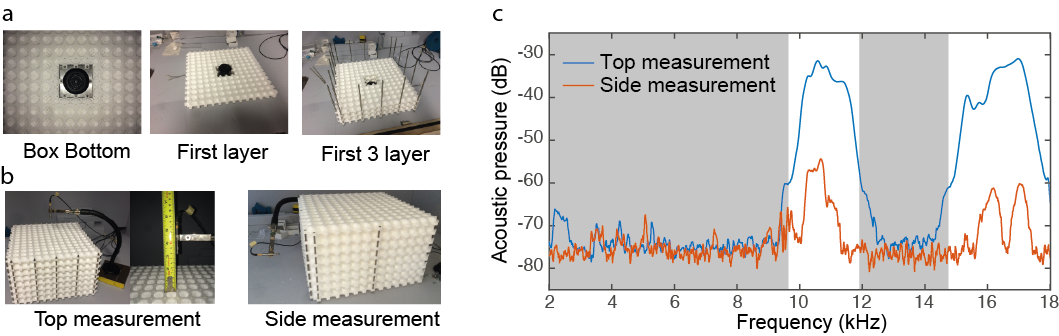}
	\end{center}
	\linespread{1.5}\caption{(a) A metamaterial assembly of 13 x 13 x 3 unit cells with a speaker attached to the bottom layer as a source for airborne sound. (b) Fully assembled metamaterial box with microphones positioned facing the speaker (left) and to its side (right). (c) The measured acoustic pressure through the metamaterial acquired through forward-facing and side-facing microphones.}
	\label{fig:acoustic_side}
\end{figure}

\paragraph{Three-dimensional metamaterial fabrication:} To characterize the metamaterial in all directions, we fabricate a ``box" consisting of 13 x 13 x 8 unit cells 25 mm each. To speed up the printing process, simplify the removal of the excess printing-powder and ease the mounting of vibrations and noise sources, we print each of the box layers separately. Five of the printed layers are composed of 13 x 13 unit cells, while the remaining three layers have a void with an equivalent space of 3 x 3 unit cells. The void hosts both a loud speaker and a piezoelectric transducer. To insure the alignment of the unit cells in the printed layers, we incorporate 13 holes at each side of printed layers. Following the layers assembly, a long screw passes vertically through the holes and is secured with two bolts at each of its ends (Fig.\ref{fig:acoustic_side}a).

\begin{figure}
	\begin{center}
		\includegraphics[width = \textwidth]{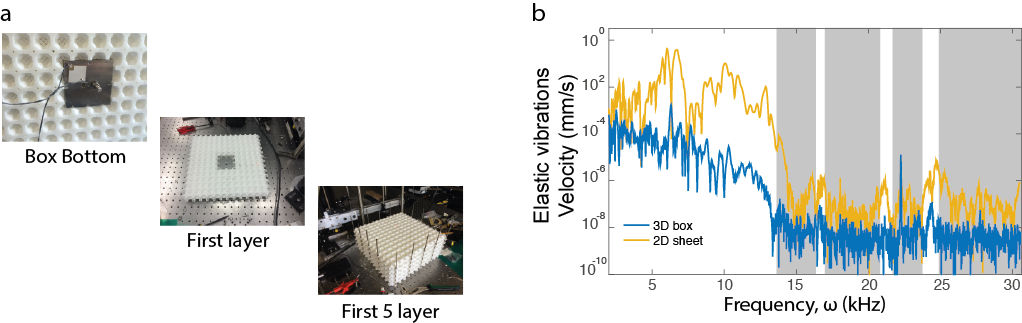}
	\end{center}
	\linespread{1.5}\caption{(a) A metamaterial assembly of 13 x 13 x 5 unit cells with a piezoelectric plate attached to the bottom layer as a source for mechanical vibrations. (b) The measured velocity transmitted through one sheet of the metamaterial at the 5th unit cell and the measured response after assembling 5 layers of the metamaterials in the vertical direction.}
	\label{fig:Box_vs_sheet}
\end{figure}

\paragraph{Three-dimensional metamaterial characterization:} To test the response of the metamaterial to airborne sound, we excite the box from within with a loud speaker (Fig.\ref{fig:acoustic_side}a). Since the box is three-dimensional, there is no need for covering the pores of the metamaterial. The box is tested within the same acoustic chamber to insure insulation from surrounding acoustic noise. We place the microphone at each one of the sides of the metamaterial box to capture the acoustic radiation in all directions (Fig.\ref{fig:acoustic_side}b). Both the side and top measurements confirm the existence of the band gaps predicted from the unit cell analysis (gray regions in (Fig.\ref{fig:acoustic_side}c)). It should be noted, however, that the transmission in the side measurements has generally lower amplitude than the top one, as the speaker is facing upwards.

To test the response of the metamaterial to elastic vibrations, we excite the box from within using a piezoelectric plate (Piezo Systems  25 x 25 x 2 mm)[www.Piezo.com]. The transmitted vibrations are then measured using the laser Doppler vibrometer at various points within the metamaterial. The measurements taken for a single sheet (2D) agree well with the full box measurement (Fig.\ref{fig:Box_vs_sheet}b).

\paragraph{Load baring capacity of the metamaterial:} 

\begin{figure}
 	\begin{center}
 		\includegraphics[scale =1]{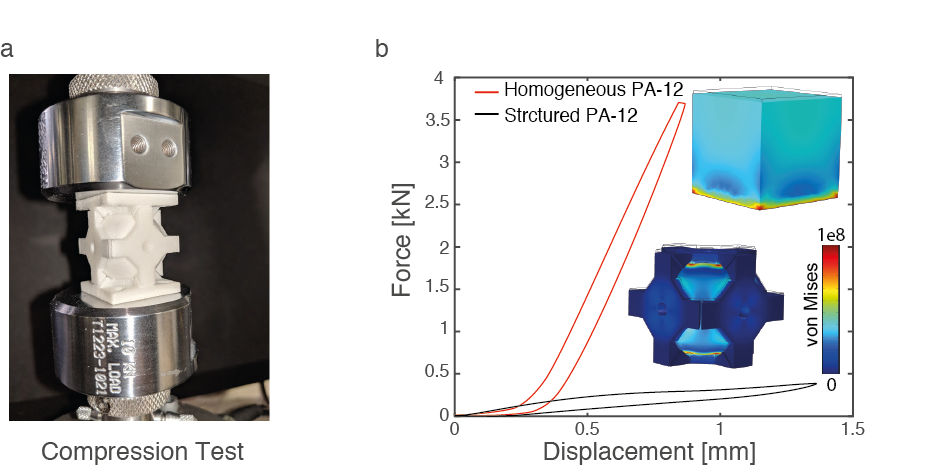}
 	\end{center}
 	\linespread{1.5}\caption {Experimental characterization of the load bearing capabilities of the structured metamaterial in comparison to a homogeneous cube of the same material (PA-12). The inset shows the numerical calculation of the Von Mises stress for a homogeneous and structured cube under compression load with 1 mm strain.}
 	\label{fig:load_comp}
 \end{figure}

We characterize the effective static stiffness of the metamaterial by comparing a block of PA-12 against a single unit cell in a compression test using an Instron 3000 machine. As expected,  the thin features of the unit cell causes a reduction in stiffness of the metamaterial by $\approx$ an order of magnitude in reference to the bulk material. A complementary numerical simulation of the compression test shows the stress concentration within the unit cell (inset in Fig. \ref{fig:load_comp}). We use COMSOL structure mechanics module to perform the numerical test. We add a prescribed displacement as a boundary condition on the top surface of the unit cell in the -z direction, while keeping the bottom face of the unit cell fixed. The numerical results suggest that the stiffness of the metamaterial can be greatly increased by increasing the thickness of the connecting shells within the unit cell. It is worth noting that the metamaterial is lighter than the bulk PA-12 by a factor 6, due to material removal.
 
\end{spacing}

\end{document}